\documentstyle[preprint,aps]{revtex}

\begin{document}
\draft

\date{\today}
\title{
Quantum Molecular Dynamics Approach to the Nuclear Matter 
\\
Below the Saturation Density
}

\author{
Toshiki~Maruyama$^1$, Koji~Niita$^{1,2}$, Kazuhiro~Oyamatsu$^3$
Tomoyuki~Maruyama$^1$, Satoshi~Chiba$^1$, and Akira~Iwamoto$^1$,
\\[1ex]
$^1$ Advanced Science Research Center, \\
Japan Atomic Energy Research Institute, \\
Tokai, Ibaraki, 319-11 Japan \\[1ex]
$^2$ Research Organization for Information Science and Technology, \\
Tokai, Ibaraki, 319-11 Japan\\[1ex]
$^3$ Department of Energy Engineering and Science, Nagoya University\\
Furo-cho, Chikusa-ku, Nagoya 464-01, Japan
}

\maketitle

\begin{abstract}
Quantum molecular dynamics is applied to study 
the ground state properties of nuclear matter at subsaturation densities.
Clustering effects are observed as to soften the 
equation of state at these densities.
The structure of nuclear matter at subsaturation density  shows 
some exotic shapes with variation of the density.
\end{abstract}


%------------------------------------------------------------------------
\section{Introduction}
%------------------------------------------------------------------------

One of the main interests of heavy-ion physics and astrophysics 
is the property of nuclear matter in extreme conditions.
Its high-density behavior is important for the scenario of supernova
explosions, the evolution of neutron stars,
the reaction process of high-energy heavy-ion collisions,
quark-gluon plasma and so on.
Properties of the nuclear matter below saturation density,
on the other hand, are essential in describing the multi-fragmentation 
in the heavy-ion collisions, the collapsing stages in supernova 
explosions and the structure of neutron star crusts.
Here the saturation density is 
the density of 
the energy-minimum state
of the nuclear matter at a fixed proton ratio.
For symmetric nuclear matter, the saturation density is the
normal nuclear density $\rho_0=0.165\;{\rm fm}^{-3}$.
Since the matter is unstable below the saturation density, 
an inhomogeneous state is expected to appear
below the saturation density.

Besides, supernova matter (SNM) and neutron star matter (NSM) are also 
interesting from the viewpoint of the nuclear shape. 
At low densities, nuclei in these matters are expected to be
crystalized so as to minimize the long range Coulomb energies.
They melt into uniform matter at a certain density close 
to the saturation density.
Then what happens in between?
About a decade ago, three groups \cite{Rav83,Has84,Wil85}
suggested that nuclei have exotic structures 
in the SNM and/or NSM.
They showed that the stable nuclear shape changes from sphere to 
cylinder, slab, cylindrical hole, 
and to spherical hole with increase of the matter density. 
The favorable nuclear shape is determined by a balance between the 
surface and Coulomb energies.
In the liquid drop model, a simple geometrical argument
demonstrates that the favorable shape changes as a function of 
the volume fraction of the nucleus in the cell, 
independently of specific nuclear interactions \cite{Has84,Oya84}.
>From recent studies in the liquid drop model \cite{Lor93} and 
the Thomas Fermi calculations \cite{Oya93,Sum95}, 
the non-spherical shapes of nuclei are expected in a density 
range at about half the saturation density although the range
depends on the choice of nuclear interaction.
Furthermore these shapes are expected to survive even if the shell 
effects are taken into account \cite{Oya94}.

These exotic nuclear shapes may cause substantial astrophysical consequences. 
In SNM, neutrino absorption by nuclei may 
modify leptonic energy of the matter and lead to a significant 
change in the strength of supernova explosions \cite{Lor93}.
In neutron stars, the exotic nuclei may also affect the pinning of 
superfluid neutron vortices to nuclei in inner crusts of neutron stars, 
which are considered to be the initial step of neutron star glitches.
Recently, Mochizuki and Izuyama \cite{MI95} demonstrated that the exotic
nuclear shapes actually play an essential role in trapping vortices in a
microscopic mechanism of the glitches.

The spatial fluctuation is important in describing properties of SNM 
and NSM because geometrical distributions of nucleons may affect
neutrino reaction rate in SNM and interactions with vortices in NSM. 
However, there have been only two works \cite{Wil85,Las87} which allow
arbitrary nuclear shapes in their models. 
These two works \cite{Wil85,Las87} use 
the Thomas Fermi approximation and treat SNM
while none has been done for NSM.
Williams and Koonin \cite{Wil85} have investigated 
the structure of neutron-proton symmetric nuclear matter 
with the proton ratio $Z/A=0.5$, 
while the authors of Ref.~\cite{Las87} 
studied the matter with $Z/A=0.285$ at the total 
entropy per baryon $s=1$.
Unfortunately, some spatial fluctuation due to a possible 
cluster correlation or non uniform distribution is neglected 
since their calculations with the Thomas-Fermi approximation 
are based on a one-body treatment of the matter.

In the field of nuclear reaction study, molecular dynamics has 
become one of the most powerful approach to simulate the fragmentation.
Advantages of molecular dynamics for the investigation of
heavy-ion reactions are that no reaction mechanism is required to assume
and that the fluctuation of the system is automatically included.

Within various kinds of molecular dynamics,
the quantum molecular dynamics (QMD)
\cite{Aic86,Aic91,Pei91,Pei92,Mar92PTP,Nii95}
approach has been proposed to study high energy heavy-ion collisions.
QMD has been also used for the analysis of fusion reaction, nucleon-induced 
reaction, fragmentation in collisions between heavy systems, and so on.

In this paper, we 
apply the QMD method to the investigation of the 
equation of state (EOS) and 
the structure of nuclear matter at subsaturation densities.
A similar calculation using QMD 
is done in Ref.~\cite{Pei91}
aiming to see the EOS of the nuclear matter.
It was reported that the clustering in the matter
significantly softens the EOS below saturation density. 
They have used, however, only 254 nucleons in 
a cell to simulate finite-temperature nuclear matter.
This number seems insufficient to investigate the nuclear matter properties.
We then perform the QMD
calculation by using much more particles in a cell,
and see the EOS of nuclear matter at zero-temperature.

In the present QMD,
we introduce the Pauli potential and 
the momentum-dependent interaction in order to 
simulate the Fermionic feature in a phenomenological way
and the energy-dependence of the optical potential.
The binding energies of finite nuclei
and the saturation properties  
of the nuclear matter are well adjusted to the empirical values. 
As a result of this new version of the QMD, 
we can approximately describe the properties of 
neutron-rich isotopes such as the density profile of $^{11}{\rm Li}$ and 
the effects of surface neutrons in the low-energy collisions \cite{refNRich}.
With these ingredients of the QMD and 
the periodic boundary condition, 
we investigate the ground state properties of the nuclear matter
at subsaturation densities.

In Sect.~\ref{secQMD} we describe our model based on the QMD.
In Sect.~\ref{secEOS}, the EOS and the structure of  nuclear matter 
is discussed.
Finally, summary is given in Sect.~\ref{secSummary}.

%------------------------------------------------------------------------
\section{QMD simulation for nuclear matter}\label{secQMD}
%------------------------------------------------------------------------

The present QMD code 
is an improved version of our 
previous one \cite{Nii95}.
In the previous works \cite{Nii95,Chi},
where we have employed very simple effective interactions in the QMD,
we have nicely reproduced several
observables in the nucleon-induced reactions 
at energy region of about 100 MeV$\sim$5 GeV.
Several improvements of this QMD are done
to enable the calculations for 
the ground state of 
the nuclear matter 
for wide range of
density and proton ratio. 
First we have introduced a phenomenological potential,
namely Pauli potential to describe the Fermionic feature of nucleons.
By this Pauli potential, we can uniquely determine the 
ground state of the finite nuclei and the nuclear matter
by seeking the energy minimum state.
Second we have introduced momentum-dependent interaction,
which is also an important feature of the Fermion system 
with finite range interaction.
In the following, we explain the detail of the present QMD 
and how it describes the nuclear matter.

%------------------------------------------------------------------------
\subsection{The total wave-function and the equation of motion}
%------------------------------------------------------------------------

In QMD, each nucleon state is represented by a Gaussian 
wave-function of width $L$,
\begin{equation}
\phi_i({\bf r}) = \frac{1}{(2\pi L)^{3/4}} \exp \left[
                - \frac{({\bf r} - {\bf R}_i)^2}{4L} +
                  \frac{i}{\hbar} {\bf r} \cdot {\bf P}_i \right],
\end{equation}
where ${\bf R}_i$ and ${\bf P}_i$ are the centers of position
and momentum of $i$-th nucleon, respectively.
The total wave-function is assumed to be a direct product of
these wave-functions.
Thus the one-body distribution function is obtained by the Wigner
transform of the wave-function,
\begin{equation}
f({\bf r},{\bf p}) =  \sum_i { f_i({\bf r},{\bf p}) },
\label{f0}
\end{equation}
\begin{equation}
f_i({\bf r},{\bf p})  =  8 \cdot
                        \exp\left[-{({\bf r}-{\bf R}_i)^2\over 2L}
                      -{2L({\bf p}-{\bf P}_i)^2\over \hbar^2}\right].
\label{fi}
\end{equation}

The equation of motion of ${\bf R}_i$ and ${\bf P}_i$ is given
by the Newtonian equation 
\begin{equation}
\dot{{\bf R}}_i =   \frac{\partial H}{\partial {\bf P}_i},
\;\;\;\;
\dot{{\bf P}}_i = - \frac{\partial H}{\partial {\bf R}_i},
\label{newton00}
\end{equation}
and the stochastic N-N collision term.
Hamiltonian $H$ consists of the kinetic energy 
and the energy of the two-body effective interactions.

%------------------------------------------------------------------------
\subsection{Effective interactions}\label{secEffectiveInt}
%------------------------------------------------------------------------

The Hamiltonian is separated into several parts as follows,
\begin{equation}
H  =  T + V_{\rm Pauli} + V_{\rm local} + V_{\rm MD}\;,
\label{ham0}
\end{equation}
where $T$, $V_{\rm Pauli}$, $V_{\rm local}$ and $V_{\rm MD}$ 
are the kinetic energy, the Pauli potential, 
the local (momentum-independent) potential
and the momentum-dependent potential parts, respectively.

The Pauli potential \cite{Pei92,Boa88,Ohn92,Mar95} is introduced 
for the sake of simulating Fermionic properties in a semiclassical way.
This phenomenological potential prohibits nucleons of the same spin $\sigma$ 
and isospin $\tau$ from coming close to each other in the phase space.
Here we employ the Gaussian form of the Pauli potential \cite{Pei92} as
\begin{equation}
  V_{\rm Pauli} = 
  \frac{1}{2} 
  C_{\rm P}\left( \frac{\hbar}{q_0 p_0}\right)^3
  \sum_{i, j(\neq i)} 
  \exp{ \left [ -\frac{({\bf R}_i-{\bf R}_j)^2}{2q_0^2} 
          -\frac{({\bf P}_i-{\bf P}_j)^2}{2p_0^2} \right ] }
  \delta_{\tau_i \tau_j} \delta_{\sigma_i \sigma_j}\;.
  \label{Pauli}
\end{equation}

In the local potential part we adopt the Skyrme type with
the Coulomb and the symmetry terms as explained 
in Eq.~(5) of Ref.~\cite{Nii95}, 
\begin{eqnarray}
V_{\rm local} & = & \; {\alpha\over 2\rho_0}\sum_i<\rho_i>
        \; + \; {\beta\over (1+\tau)\;\rho_0^{\tau}}
                \sum_i<\rho_i>^{\tau} \nonumber \\
  &   & \; + \; {e^2 \over 2}\sum_{i , j(\neq i)} c_{i} \, c_{j}
          \int\!\!\!\!\int d^3r_i\,d^3r_j 
                 { 1 \over|{\bf r}_i-{\bf r}_j|} \,
                \rho_i({\bf r}_i)\rho_j({\bf r}_j)
                \nonumber \\
  &   & \; + \; {C_{\rm s}\over 2\rho_0} \sum_{i , j(\neq i)} \,
                ( 1 - 2 | c_i - c_j | ) \; \rho_{ij}.
\label{ham1}
\end{eqnarray}
In the above equation, $c_i$ is 1 for protons and 0 for neutrons,
while $<\rho_i>$  is an overlap of density with
other nucleons defined as
\begin{eqnarray}
<\rho_i>     & \equiv & \sum_{j(\neq i)} \; \rho_{ij} \;
              \equiv  \sum_{j(\neq i)}
                     { \int { d^3r \; \rho_i({\bf r}) \;
                       \rho_j({\bf r}) }} \nonumber \\
             & = & \sum_{j(\neq i)}{ (4\pi L)^{-3/2}
                  \exp \left[ - ( {\bf R}_i - {\bf R}_j ) ^2
                  / 4L \right] }\;.
\label{rhoij}
\end{eqnarray}

The momentum-dependent term is introduced as a Fock term of 
the Yukawa-type interaction.
We divide this interaction into two ranges
so as to fit 
the effective mass and the energy dependence of the 
real part of the optical potential,
as
\begin{eqnarray}
V_{\rm MD}  & = & V_{\rm MD}^{(1)} + V_{\rm MD}^{(2)} \nonumber \\
 & = &
         {C_{\rm ex}^{(1)} \over 2\rho_0} \sum_{i , j(\neq i)} 
      {1 \over 1+\left[{{\bf P}_i-{\bf P}_j \over \mu_1}\right]^2} 
      \;\rho_{ij}
  +   {C_{\rm ex}^{(2)} \over 2\rho_0} \sum_{i , j(\neq i)} 
      {1 \over 1+\left[{{\bf P}_i-{\bf P}_j \over \mu_2}\right]^2} 
      \;\rho_{ij}\ .
\label{eqMD}
\end{eqnarray}

The parametrization of the constants included in the above 
effective interactions will be discussed in Sec.~\ref{secParameters}.

%------------------------------------------------------------------------
\subsection{Energy minimum state}
%------------------------------------------------------------------------

With inclusion of the Pauli potential, 
we can define the ground state
as an energy-minimum state of the system. 
To get the energy-minimum configuration, 
we use the following damping equation of motion,
\begin{equation}
\dot{{\bf R}}_i =   \frac{\partial H}{\partial {\bf P}_i}
       +\mu_{\bf R} \frac{\partial H}{\partial {\bf R}_i},
\;\;\;\;
\dot{{\bf P}}_i = - \frac{\partial H}{\partial {\bf R}_i}
       +\mu_{\bf P}\frac{\partial H}{\partial {\bf P}_i},
\label{damp00}
\end{equation}
where $\mu_{\bf R}$ and $\mu_{\bf P}$ are the damping coefficients
with negative values when we need to cool the system.

We first distribute the particles randomly in the phase space
and cool down the system according to the damping equation of motion 
until the energy reaches the minimum value.
Sometimes the system stops at the local minimum during the cooling.
We thus try again and again this cooling procedure
with a different initial state and seek the real energy minimum state.

For finite nucleus and infinite system 
above saturation density,
this procedure works rather well.
For infinite system at subsaturation densities, however,
there are many local minimum states around the real ground state, 
which differ from the ground state
in the detail of the surface configuration of the clusters.
Since the energy difference from the ground state 
is the order of 10 keV/nucleon 
in this case, we accept these states as ground states 
and neglect the small differences of the configuration.

%------------------------------------------------------------------------
\subsection{Periodic boundary condition}
%------------------------------------------------------------------------

In order to simulate infinite nuclear matter with finite number of particles,
we use a cubic cell with periodic boundary condition.
The size of the cell is determined from the average density
and the particle number.
The periodic boundary condition is introduced as follows:
As shown in Fig.~\ref{figCell}, we prepare 26 ($=3^3-1$) 
surrounding cells, where the particle distribution 
reflects the distribution of the central cell exactly.
The particles in the central cell move according to the interaction 
with all particles in the same cell and in the surrounding cells
as well.
The particles in the surrounding cells obey exactly the same motions as 
those in the central cell.
Thus the Hamiltonian per cell is written as
\begin{equation}
H=\sum_{i=1,\cdots, N}
\bigg[\;\; T_i
%  +\sum_{{\rm cell}=0,\cdots, 26 \ } \sum_{j=1,\cdots, N}
  +\sum_{{{\rm cell}=0,\cdots, 26 \ }
          \atop {j=1,\cdots, N}}
     H^{(2)}_{ij}({\bf R}_i-{\bf R}_j+{\bf D}_{\rm cell},\ {\bf P}_i, {\bf P}_j)
  +\cdots\ \ 
\bigg]\ ,
\label{Hamiltonian}
\end{equation}
where $T_i$ is one-body part (kinetic energy), $H^{(2)}_{ij}$ is the 
two body part of the Hamiltonian and ${\bf D}_{\rm cell}$ are 
the relative position of surrounding cells from the center.
Note that the indices ``cell'' runs from 0 (the center cell) to 26 
(surrounding cells) and ${\bf D}_0=0$.

%------------------------------------------------------------------------
\subsection{Parametrization of the constants}\label{secParameters}
%------------------------------------------------------------------------

We have twelve parameters in the effective interactions
of the Hamiltonian Eq.~(\ref{ham0}), i.e.,
$C_{\rm P}, q_0, p_0$, $\alpha, \beta, \tau, C_{\rm s}$, 
$C^{(1)}_{\rm ex}, C^{(2)}_{\rm ex}, \mu_1, \mu_2$
and the Gaussian width $L$. 
We should parametrize these constants 
to reproduce properties of the ground states of the finite nuclei and
saturation properties of the nuclear matter.

We first determine the parameters of Pauli potential
$q_0, p_0$ and $C_{\rm P}$, apart from the other effective interactions,
by fitting the kinetic energy of the exact Fermi gas
at zero temperature and at various densities.
For this, we define the free Fermi gas system as a ground state 
for the Hamiltonian including only 
the kinetic energy and the Pauli potential 
by making use of the damping equation of motion Eq.~(\ref{damp00}) 
and the periodic boundary condition with 1024 particles in a cell.
In Fig.~\ref{figPauli}, we show the kinetic energies 
(the solid squares) and the total energies (the open squares)
obtained by using a parameter set for the Pauli potential as
\begin{equation}
  C_{\rm P}=207\;{\rm MeV},\;\;\; 
  p_0=120\;{\rm MeV}/c,\;\;\;
  q_0=1.644\;{\rm fm}.
\end{equation}
In the same figure we draw the exact energy of 
the Fermi gas by the solid line.
Although there are some other parameter sets 
which can reproduce the exact energies of the Fermi gas
in the same form,
i.g. that used in Ref.~\cite{Pei91},
we choose the above parameter set to get good properties 
of the ground states of the finite nuclei
with other effective interactions particularly 
in combination with 
the momentum-dependent interaction.

Among remaining nine conditions, 
four are attributed to the momentum-dependent interaction
as follows.
%In order to determine the four parameters 
%$C_{\rm ex}^{(1)}$, $C_{\rm ex}^{(2)}$, $\mu_1$, and $\mu_2$,
We calculate the single particle potential of momentum ${\bf p}$
in ideal nuclear matter at the normal nuclear density,
which leads to
\begin{eqnarray}
U({\bf p}, \rho_0) & = & U_{\rm local} + U_{\!\rm MD}({\bf p})
\nonumber  \\
 & = & \alpha  + \beta 
  + \left( \frac{4}{3}\pi p_{\rm F}^3 \right)^{-1}
    \int^{p_{\rm F}} d^3 p'
    \left[{C_{\rm ex}^{(1)} \over 1
  + \left[ {{\bf p}-{\bf p'} \over \mu_1} \right]^2} 
  + {C_{\rm ex}^{(2)} \over 1+\left[ {{\bf p}-{\bf p'} \over \mu_2}
\right]^2} \right] 
\nonumber  \\
 & = & \alpha  + \beta 
    + C_{\rm ex}^{(1)} g(x=\mu_1/p_{\rm F}, y=p/p_{\rm F})
    + C_{\rm ex}^{(2)} g(x=\mu_2/p_{\rm F}, y=p/p_{\rm F}),
\label{eqUOpt}
\end{eqnarray}
with
\begin{equation}
g(x,y) = \frac{3}{4} x^3 \left[ 
         \frac{1+x^2-y^2}{2xy}
         \ln{ \frac{(y+1)^2+x^2}{(y-1)^2+x^2}}
       + \frac{2}{x}
       - 2 \left\{ \arctan{ \frac{y+1}{x} }
                - \arctan{ \frac{y-1}{x} }
           \right\} \right] .
\end{equation}
We fit the energy-dependence of
this potential to the experimental data.
In Fig.~\ref{figPdep}, we plot the energy dependence of the real
part of the optical potential (the open circles and squares)
obtained from the experimental data 
of Hama et al.~\cite{Ham90} for p-nucleus elastic scattering.
>From this figure, we pick up three conditions, i.e. 
$U(0) = -80$ MeV, $U(p)=0$ at $E_{\rm lab} = 200$ MeV, and 
$U(p \rightarrow \infty) = \alpha + \beta = 77$ MeV.
For another condition, we use the value of effective mass defined by
\begin{equation}
 \frac{1}{m^*} = \frac{1}{m} +
 \left( \frac{1}{p} 
 \frac{\partial U_{\!\rm MD}}{\partial p} 
 \right)_{p=p_{\rm F}}.
\end{equation}
We take  the value of $m^*=0.8\,m$ at $\rho=\rho_0$.

Other three conditions are coming from the saturation condition,
i.e. the energy per nucleon $E/A=-16$ MeV 
at $\rho=\rho_0$ (0.165 fm$^{-3}$) 
and the value of incompressibility $K$.

The last two parameters are given by hand.
One is the value of the coefficient of the symmetry term
$C_{\rm s}$.
We take 25 MeV for $C_{\rm s}$.
This value leads to the symmetry energy 34.6 MeV 
for the nuclear matter at the saturation density
(see Sec.~\ref{subsecAsy}).
The other is the width of the Gaussian wave packet $L$,
which is a free parameter in QMD model.
This value affects ground state properties of finite nuclei
and infinite nuclear matter below saturation densities,
while it does not change those of infinite nuclear matter above
saturation densities.
We then choose this value to give nice fitting the binding 
energies of finite nuclei.

It should be noted here that
we cannot determine these parameters from the above conditions 
in an analytical way, since the Fermi distribution is not 
exactly achieved by the Pauli potential and 
the additional potential energy included in the Pauli potential.
In addition, 
the saturation properties of the nuclear matter 
should be realized in the simulated matter
for the main purpose of this paper.
We then simulate the nuclear matter by the QMD with 
the periodic boundary condition using 1024 particles in a cell. 
We search the energy minimum state
by the damping equation of motion as discussed above
and adjust the parameters.
By this method, we have fixed three parameter sets 
according to the value of incompressibility $K$, 
which are shown in Table 1. 
We have prepared 
three kinds of equation of state, namely Soft ($K$=210 MeV),
Medium ($K$=280 MeV) and Hard ($K$=380 MeV) EOS.
These values of incompressibility $K$ are subtracted 
from the results of EOS (shown in Fig.~\ref{figRhoE})
by fitting its curvature at the 
saturation point to the following parabolic form,
\begin{equation}
 E/A=\frac{K}{18\rho_0^2}\left(\rho-\rho_0\right)^2-16\;.
\end{equation}
This parabola is also shown in Fig.~\ref{figRhoE}.

Here the single particle potential shown in Fig.~\ref{figPdep}
are also calculated by the simulated nuclear matter 
with Pauli potential and other effective interactions 
instead of ideal nuclear matter.
The results are denoted by the crosses in Fig.~\ref{figPdep}
and well coincident with the results of ideal nuclear matter
except for the low energy part, where the Pauli potential is 
effective.
Though this result in Fig.~\ref{figPdep}
is obtained with a parameter set of Medium EOS, 
results with Soft and Hard EOS are same as 
Medium EOS within 2 MeV for all energy region.

In Fig.~\ref{figEbind}, we plot 
the binding energies of the ground state of finite nuclei 
obtained by the damping equation of motion Eq.~(\ref{damp00})
with three parameter sets, i.e., Soft (the long dashed line), 
Medium (the dashed line) and Hard (the solid line) EOS.
All of them reproduce well the global trend of the binding energies 
of various nuclei except for light nuclei from 
$^{12}$C to $^{20}$Ne.
It might be due to
the specific structures of these light nuclei,
which are not well 
described by the present QMD.

\subsection{``Screened'' Coulomb potential}

For the neutron star, the same number of electrons exist as proton, 
since the nuclear matter in the neutron star should be charge-neutral. 
Hence the energy of the system remains finite
even if we calculate the Coulomb interaction of
protons and electrons.
However, Coulomb interaction has so long range that
the Coulomb energy depends on 
the cell-size 
in our treatment of the infinite system by the periodic 
boundary condition with the surrounding neighbor cells.
To avoid this cell-size dependence, we introduce
a cutoff of the Coulomb interaction in the way of
``screened'' Coulomb potential.
We use the following  ``screened'' Coulomb interaction
instead of the second term of Eq.~(\ref{ham1})
for the nuclear matter calculations,
\begin{equation}
      V_{\rm C}^{\rm scr} =
    {e^2 \over 2}\sum_{i , j(\neq i)} c_{i} \, c_{j}
     \int\!\!\!\int d^3r_i\,d^3r_j { 
    \exp\left[-|{\bf r}_i-{\bf r}_j|/r_{\rm scr} \right] 
              \over|{\bf r}_i-{\bf r}_j|} \,
           \rho_i({\bf r}_i)\rho_j({\bf r}_j)\;.
\end{equation}
In this equation, $r_{\rm scr}$ is the ``screening'' length,
which we use 10 fm in the present study.
The physical screening of the Coulomb potential by the electron
localization is, however, estimated to be much larger in the case of 
normal nuclear density \cite{Wil85}.
%%Our ``screening'' length to include is much shorter and
%%has no physical meaning except for avoiding the cell-size dependence.
Thus our ``screening'' should be considered as a technical approximation to
avoid this cell-size dependence and to
make the numerical calculation feasible.
For this purpose, $r_{\rm scr}$ should be smaller than the cell size.
On the other hand, to keep the proper description of finite nuclei,
it should be lager than the size of nuclei.
By the ``screened'' Coulomb interaction with $r_{\rm scr}$ = 10 fm,
however, the binding energies 
are slightly modified particularly 
in heavy nuclei.
We compare the binding energies obtained by the ``screened''
Coulomb interaction with that of the normal one in Fig.~\ref{figEbind}.
The dotted line is the result of the ``screened'' Coulomb 
interaction with Medium EOS.
Though the binding energies of heavy nuclei increase,
the binding energies of finite nuclei still have 
the maximum around at $A\approx 100$.
This feature is important 
to describe clustering of the matter at low densities.

%------------------------------------------------------------------------
\section{Equation of state and the structure of nuclear matter}
%------------------------------------------------------------------------
\label{secEOS}

In this section we study properties of
nuclear matter at several conditions.
It is desirable to use a cell large enough to include 
several periods of structure and
to avoid the spurious effects of boundary condition on the
structure of the matter.
Though our calculation with typically 1024 particles in a cell
is not fully satisfactory in this respect,
we consider it is enough for semiqualitative discussions 
at the beginning of this study.
Actually, the global quantities, i.g.~ the ground state energy of the system, 
%% which is obtained by the damping equation of motion,
is well saturated at this number of particles in a cell.
In Fig.~\ref{figNdep}, we show the energy per nucleon  
of the infinite system as a function of the particle number 
in a cell for four average densities from $\rho=0.4 \rho_0$ 
to $2.2 \rho_0$.
For all densities, the energy of the system have already approached 
the asymptotic value above 256 particles within 100 keV/nucleon.
In this study we then simulate the infinite system 
by the periodic boundary condition 
with 1024 or 2048 particles in a cell, and investigate  
ground state properties of the nuclear matter.

%------------------------------------------------------------------------
\subsection{Symmetric nuclear matter}
%------------------------------------------------------------------------

We first perform the calculations for symmetric 
($Z/A=0.5$) matter at zero temperature to simulate supernova 
matter (SNM) in the collapsing stage.
Figure \ref{figRhoE} shows 
the energy per nucleon
as a function of the average density.
The solid squares indicate the energy of ``uniform'' nuclear matter
while open squares are the results of energy-minimum configurations.

The ``uniform'' matter energy is calculated as follows:
First we distribute nucleons randomly and cool the system
only with the Pauli potential.
Pauli potential is repulsive and does not 
spoil the uniformity of the system.
Then we impose the other effective interactions
and cool only in the 
momentum space fixing the positions of particles.
The system turns out to be approximately uniform with this procedure.
Note that the simulated ``uniform'' matter is not exactly the same as ideal 
nuclear matter since the latter is continuous and completely uniform.

Both cases of uniform and energy-minimum configurations
have almost the same energy per nucleon for the higher densities
as is seen in this figure.
Below saturation density $\rho_0$, the energy per nucleon of
the energy-minimum configuration is lower than the uniform case.
The deviation amounts to about 5 MeV.
As we see in the following, this is due to the structure change of the
matter from uniform to non uniform structure such as clusterized one.

This change of the structure is displayed in Fig.~\ref{figMattSym}.
Above 0.8 $\rho_0$, the system is almost uniform and no specific
structure is seen.
Below 0.8 $\rho_0$, however, there appear some voids between the matter.
As the density gets lower, the voids develop and 
nuclei are surrounded by the voids.
Below 0.2 $\rho_0$ each nucleus is separate, while
above that density nuclei are connected to form 
some transient structures.
This change of the structure is basically the same as reported 
in the previous works \cite{Rav83,Has84,Wil85}.
Furthermore, the transient shape of the matter like
hole, slab and cylinder are partly seen in our calculation although
the nuclear surface shape in QMD is somewhat complicated.
It should be noted that the nuclear shapes do not show exact symmetry
properties (sphere, cylinder or slab) assumed in the previous liquid-drop
and Thomas-Fermi studies \cite{Rav83,Has84,Oya84,Lor93,Oya93,Sum95},
where the assumption of the symmetries leads to
clear changes of the nuclear shape.
Our result suggests that during the transition from 
homogeneous to inhomogeneous matter, the nuclear shape 
may not have these simple symmetry properties. 
It is possible that this is due to the incomplete
minimization of the energy 
because 
there are several local minimums around the real ground state and
the energy difference due to the nuclear shape is extremely small. 
However, 
these shapes are not so strange 
in the transient SNM at the collapsing stage because the matter
is not in perfect equilibrium.

The change of the shape obtained in the QMD calculation may affect 
the neutrino reaction rate in SNM. 
It has been pointed out \cite{Rav83,Has84,Wil85} that 
the nuclear distributions are essentially determined 
by a delicate balance
(of the order of 1 keV/nucleon) 
between the surface and Coulomb energies.
In the present treatment, however, we neglect such a tiny energy 
difference between the nuclear shapes around the ground state.
We then need further improvement of our treatment to investigate 
the affect of the shape on the neutrino reaction rate in SNM.

Nevertheless, our results provide the EOS of the symmetric matter with
sufficient accuracy 
as well as global nuclear structure in the matter.

%------------------------------------------------------------------------
\subsection{Asymmetric nuclear matter}\label{subsecAsy}
%------------------------------------------------------------------------

In the neutron star matter (NSM), the beta-equilibrium is achieved 
and the proton ratio is given by the energy-minimum condition.
The left panel of Fig.~\ref{figZE} is the energy per nucleon
of nuclear matter with several proton ratios.
The electron kinetic energy is not included in this figure,
though the Coulomb interaction of protons and uniform 
electron background is included.
The energy per nucleon at $\rho=\rho_0$ is fitted by
\begin{equation}
E/A\approx -16.2 + 34.6 \frac{(Z-N)^2}{A^2}\;\; {\rm [MeV]}.
\end{equation}
In other words, the symmetry energy at normal density is 34.6 MeV 
in our calculation.
The symmetry energies at $\rho=0.6\rho_0$ and $0.2\rho_0$
obtained in the same way are 23.0 and 18.9 MeV, respectively.

Including the electron kinetic energy at zero temperature, 
we get the total energy of NSM as a function of the proton ratio
shown in the right panel of Fig.~\ref{figZE}. 
It can be seen from this figure that
the proton ratio that gives the energy-minimum of the system is 
0.032 ($\rho=0.2\rho_0$),
0.043 ($\rho=0.6\rho_0$) and
0.072 ($\rho=1.0\rho_0$).
%% $0.03\sim0.08$ (it depends on the density).
%% These values agree quantitatively to the values reported 
%% in Refs.~\cite{Lor93,Oya93,Sum95}.

The structure of asymmetric nuclear matter at low density ($0.1\rho_0$)
is shown in Fig.~\ref{figMattAsym}.
Even though the proton ratio is small, nucleons form the
cluster structure at low density.
If the proton ratio is very small, some neutrons can not stay inside the 
cluster but overflow into the space; clusters are floating in 
the neutron sea.
When the proton ratio increases, free neutrons are absorbed 
into the clusters.

The departure from the spherical symmetry of the nuclear 
(cluster) shape is also seen as in the symmetric matter. 
%This may cause some consequences in the scenario of pulsar 
%glitches because the geometry of the clusters affects 
%the pinning of vortices in the neutron superfluid.
This may cause some consequences in the standard 
scenario of pulsar glitches. 
In the scenario, vortices in the superfluid neutron sea are
supposed to be pinned to clusters (nuclei) and accumulated 
in the inner crusts because neutrons are normal in nuclei. 
The strength of the pinning, which is important in the scenario, 
depends on the geometry of nuclei as well as on the 
superfluid energy gap.

Finally, we have to recognize that the simulation of infinite system 
still needs much lager number of particles.
%At least several structures must appear in a cell to get distinct conclusions.
At least a cell must include several periods of structure 
for distinct conclusions.
If only one or two units of the structure is included in a cell,
the size of the unit structure is the same or half size of the
given cell size. 
The lattice is also limited to be cubic.
In this case, the results are dependent on the boundary condition which 
is artificially imposed.
However, we emphasize that the QMD framework can be used also for
the NSM and the symmetric matter as has been used for the reaction studies.
Together with some refinements of the surface energy, 
the present method will be able 
to describe from stable and unstable nuclei to homogeneous 
and inhomogeneous nuclear matter.
The problem of the computational time in this study 
is expected to be solved soon.

%------------------------------------------------------------------------
\section{Summary}\label{secSummary}
%------------------------------------------------------------------------

We have proposed the QMD approach for the description of
nuclear matter in wide range of density and proton ratio.
We can well reproduce the finite nuclear properties
for various mass range 
%and proton ratio
by inclusion of 
the Pauli potential and the momentum-dependent interaction.
We have investigated the EOS of nuclear matter
by simulating the infinite system with our QMD.
Below the saturation density, clustering of the system
was observed, which softens the EOS by
lowering the energy per nucleon up to about 5 MeV.

We have shown the structure of the
nuclear matter at subsaturation density.
The transient shape of the symmetric nuclear matter,
such as hole, slab, cylinder and sphere, 
predicted in previous works 
with analytic model and Thomas-Fermi calculations \cite{Rav83,Has84,Wil85},
are partially seen in our calculation. 
However the structure of nuclear matter at subsaturation density appears 
rather vaguely in our case.
This result suggests that during the transition from
homogeneous to inhomogeneous matter, the nuclear shape
may not have these simple symmetry properties.
We need, however, further investigation increasing 
the particle number to get quantitative conclusion.

For asymmetric nuclear matter we have obtained 
the proton ratios $Z/A=$
0.032 ($\rho=0.2\rho_0$),
0.043 ($\rho=0.6\rho_0$) and
0.072 ($\rho=1.0\rho_0$),
which give the energy-minimum of the system
for the fixed average densities.
At considerably low proton ratio, we have observed a neutron sea
in which the normal nuclei are floating around. 

In this paper we have presented our first results on the infinite 
nuclear matter by the use of molecular dynamics method.
Though it is still necessary to enlarge the particle number,
our results agree quantitatively to previous studies
which include
much more assumptions and restrictions in the models.

Our model contains further possibility for the
simulation of dynamical evolution of infinite nuclear matter
such as supernova explosion,
the glitch of the neutron star
and the initial stage of the universe.
Intensive and systematic study of the nuclear matter
with present model will be important 
since it contains less assumption than the foregoing models as to the
structure of the matter.

%------------------------------------------------------------------------
% R E F E R E N C E S 
%------------------------------------------------------------------------

%\begin{thebibliography}{99}

%\end{thebibliography}

%------------------------------------------------------------------------
%    T A B L E 
%------------------------------------------------------------------------
\begin{table}
\caption{Effective interaction parameter set}
\begin{tabular}{cccccc}
& & Soft ($K$=210 MeV) & Medium ($K$=280 MeV) & Hard ($K$=380 MeV) &\\
\noalign{\hrule}
& $\alpha$ (MeV) & $-223.56$ & $-92.86$ & $-21.21$ \\
& $\beta$ (MeV) & 298.78 & 169.28 & 97.93 &\\
& $\tau$ & 1.16667 & 1.33333 & 1.66667 &\\
& $C_{\rm s}$ (MeV) & 25.0 & 25.0 & 25.0 &\\
& $C_{\rm ex}^{(1)}$ (MeV) & $-258.54$ & $-258.54$ & $-258.54$ &\\
& $C_{\rm ex}^{(2)}$ (MeV) & 375.6 & 375.6 & 375.6 &\\
& $\mu_1$ (MeV) & 2.35 & 2.35 & 2.35 &\\
& $\mu_2$ (MeV) & 0.4 & 0.4 & 0.4 &\\
& $L$ (fm$^2$) & 2.1 & 2.1 & 2.05 &\\
\end{tabular}
\label{table1}
\end{table}

%------------------------------------------------------------------------
%          F I G U R E S
%------------------------------------------------------------------------

\begin{figure}
\caption{
An illustrative explanation of cell configuration.
26 surrounding cells (only 8 cells are displayed in this figure)
have exactly the same distribution of particles as the central cell.
The relative position vector of each surrounding cell from the
central cell 0 is ${\bf D}_{\rm cell}$ and ${\bf D}_0=0$.
}
\label{figCell}
\end{figure}
%------------------------------------------------------------------------

\begin{figure}
\caption{
Energy per particle of free Fermi gas.
The solid line shows the exact value.
The cases of the molecular dynamics calculation
with only Pauli potential
are shown by the solid squares (kinetic energy) and
the open squares (total energy).
}
\label{figPauli}
\end{figure}
%------------------------------------------------------------------------

\begin{figure}
\caption{
The energy dependence of the real part of the optical potential.
The open circles and squares indicate the results
obtained from the experimental data 
of Hama et al.~\protect\cite{Ham90} for p-nucleus elastic scattering.
The solid line denotes the single particle potential
calculated by Eq.~(\protect\ref{eqUOpt}) in ideal nuclear matter
with the parameter set of Medium EOS.
The short and long dashed lines show 
%two components of the momentum-dependent interaction 
$\alpha  + \beta + C_{\rm ex}^{(1)} g(x, y)$
and $C_{\rm ex}^{(2)} g(x, y)$
of Eq.~(\protect\ref{eqUOpt}), respectively.
The single particle potential
calculated by the simulated nuclear matter 
with Pauli potential is shown by the crosses.
}
\label{figPdep}
\end{figure}

%------------------------------------------------------------------------

\begin{figure}
\caption{
Binding energies of finite system
obtained by the damping equation of motion
with three parameter sets, i.e., Soft (the long dashed line), 
Medium (the dashed line) and Hard (the solid line) EOS.
The solid squares denote experimental data.
The dotted line indicates the binding energy per nucleon
obtained by using the ``screened'' Coulomb interaction 
in the case of Medium EOS.
}
\label{figEbind}
\end{figure}
%------------------------------------------------------------------------

\begin{figure}
\caption{
Particle number dependence of the energy per nucleon
of the infinite system 
for four average densities from $\rho=0.4 \rho_0$ 
to $2.2 \rho_0$.
}
\label{figNdep}
\end{figure}

%------------------------------------------------------------------------

\begin{figure}
\caption{
The energy per nucleon of symmetric 
nuclear matter ($Z/A=0.5$) at zero temperature
as a function of the average densities.
>From the left, 
the open squares are the results with Soft ($K=210$ MeV),
Medium ($K=280$ MeV) and Hard EOS ($K=380$ MeV) 
obtained by the damping equation of motion searching the energy-minimum
configuration in the full phase-space.
The solid squares indicate results obtained from the spatially
uniform distribution.
The kinetic energy of the electron is not included.
We use 1024 particles in a cell for all cases.
}
\label{figRhoE}
\end{figure}

%------------------------------------------------------------------------

\begin{figure}
\caption{
The structure of symmetric nuclear matter.
>From the upper left,
the average density is $\rho = 1.0\rho_0$, $0.8\rho_0$, $0.6\rho_0$,
$0.4\rho_0$, $0.2\rho_0$ and $0.1\rho_0$.
The white circles denote neutrons and red circles are protons.
Nuclear potential of Medium EOS is used.
We use 2048 particles in a cell for these cases and
the size of a cell is indicated in the figure.
}
\label{figMattSym}
\end{figure}

%------------------------------------------------------------------------

\begin{figure}
\caption{
The left panel: 
the energy per nucleon of nuclear matter as a function of
the proton ratio $Z/A$ for the three fixed average densities
$\rho = 0.2\rho_0$ (the solid triangles), 
$0.6\rho_0$ (the solid squares), and $1.0\rho_0$ (the solid circles).
The right panel: same as the left panel but with the kinetic 
energy of the electrons.
}
\label{figZE}
\end{figure}

%%------------------------------------------------------------------------
\begin{figure}
\caption{
The structure of asymmetric nuclear matter.
>From the upper left, the proton ratio $Z/A$ is
0.098, 0.195, 0.293 and 0.391,
while the average density is $0.1\rho_0$ for all panels.
The white circles denote neutrons and red circles are protons.
Nuclear potential of Medium EOS is used.
We use 1024 particles in a cell for these cases and
the size of a cell is indicated in the figure.
}
\label{figMattAsym}
\end{figure}
%------------------------------------------------------------------------

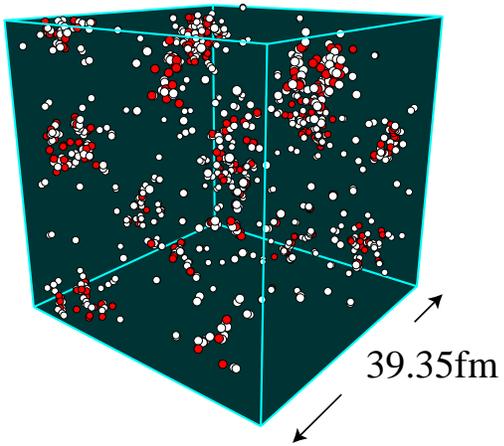
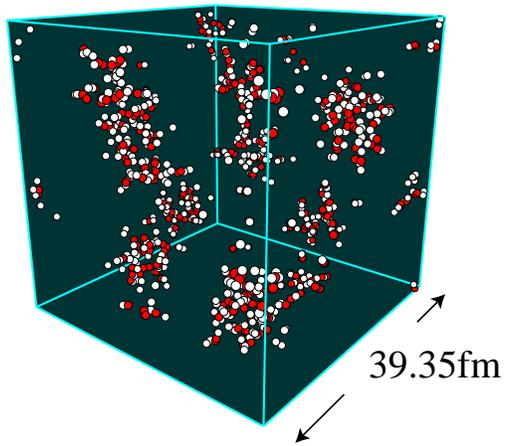
\end{document}